\documentclass[aps,prl,twocolumn,floatfix,showkeys,showpacs,groupedaddress]{revtex4}
\usepackage{latexsym} \usepackage{amsmath} \usepackage{dcolumn}
\usepackage{bm} \usepackage{graphicx} \usepackage{epsfig}

\begin{document}


\title{Finite temperature Green's function approach for excited state and thermodynamic properties of cool to warm dense matter}




\author{J. J. Kas} \affiliation{Dept.\ of Physics, Univ.\ of
Washington Seattle, WA 98195}
\author{J. J. Rehr} \affiliation{Dept.\ of Physics, Univ.\ of
Washington Seattle, WA 98195}


\date{\today}

\begin{abstract}
We present a finite-temperature extension of the retarded cumulant
Green's function for calculations of exited-state and thermodynamic
properties of electronic systems.
The method incorporates a cumulant to leading order in the
screened Coulomb interaction $W$ and improves excited
state properties compared to the $GW$
approximation of many-body perturbation theory.
Results for
the  homogeneous electron gas are presented
for a wide range of densities and temperatures,
from cool to warm dense matter regime, which
reveal several hitherto unexpected properties. For example, correlation effects remain strong at high $T$ while
the exchange-correlation energy becomes small.  
In addition, the spectral function
broadens and damping increases with temperature,
blurring the usual quasi-particle picture. Similarly Compton scattering exhibits substantial many-body corrections that 
persist at normal densities and intermediate $T$. 
Results for exchange-correlation energies and potentials are
in good agreement with existing theories and finite-temperature
DFT functionals.

\end{abstract}

\pacs{71.15.-m, 31.10.+z,71.10.-w}
\keywords {Green's function, cumulant, GW, DFT, Excited States}

\maketitle

Finite temperature (FT) effects in electronic systems are both of fundamental
interest and practical importance. These effects vary markedly depending
on whether the temperature $T$ is larger or smaller than
the Fermi temperature $T_F$ (typically a few eV).  At ``cool" temperatures,
where $T$ is much smaller than $T_F$, electrons are nearly degenerate, and
Fermi factors and excitations such as
phonons dominate
the thermal behavior \cite{Allen1982,Heine76,Draxl}.  In contrast thermal occupations become nearly
semi-classical and electronic excitations such as plasmons become important
in the warm-dense-matter (WDM) regime, where $T$ 
is of order $T_F$ or larger, and condensed matter is partially ionized.
 Recently there has been considerable interest in both experimental and
theoretical investigations of WDM for applications ranging from
laser-shocked systems
 and inertial confinement fusion to astrophysics \cite{Koenig05,Driver16}.
Many of these studies focus on equilibrium thermodynamic
properties, e.g., using the FT generalization of density
functional theory (DFT) \cite{Dharma,Duffy,Burke16}.
Although in principle, FT DFT is exact \cite{HohenbergKohn,kohnshamdft,Mermin65},
practical applications require exchange-correlation functionals
which must be approximated, e.g., by constrained fits \cite{Trickey14} to
theoretical electron gas calculations \cite{RPIMC,QMC,Burke16,Tanaka86,STLS,Perdew2000}. However, these approaches have various limitations.
First, many materials properties such as optical and x-ray spectra
depend on quasi-particle or excited-state effects.
For example, band-gaps depend on quasi-particle energies, and
calculations of x-ray spectra \cite{Prendergast11} and
Compton scattering require correlation corrections \cite{Seidler12}.
Although methods like quantum Monte-Carlo and the random phase approximation
(RPA) can provide accurate correlation energies, they are not 
directly applicable to such excited state properties. 
Secondly, currently available exchange-correlation functionals can
exhibit unphysical 
behavior outside the range of validity of theoretical data \cite{Trickey16}.
On the other hand Green's function (GF) methods within many-body
perturbation theory (MBPT) provide a systematic framework for calculations
of both excited state and thermodynamic equilibrium
properties, including total  and correlation energies.
Such methods are widely used at $T=0$, as are FT generalizations
in applications ranging from phonon-effects at low $T$ \cite{engelsbergPR63,Allen1982,Draxl,Heine76}
to nuclear matter \cite{Rios08}.  Nevertheless, while the theoretical
formalism is well established, relatively little attention has been
devoted to practical applications of FT GF methods at high $T$.
Methods for FT exchange-correlation potentials have also been
developed \cite{DharmaTaylor}, and quasi-particle corrections
have been addressed using the quasiparticle self-consistent
$GW$ approximation (QPSCGW) \cite{Schilfgaarde06}, but many 
excited state properties remain unexplored.

In an effort to address these limitations, we have developed a finite-$T$
extension of the retarded cumulant Green's function \cite{hedin99review,KasRC}. 
The cumulant approach has been successful at zero $T$ in elucidating
correlation effects beyond the $GW$ approximation of MBPT in a variety of
contexts \cite{KRC,particlehole,lischner,giustino,giustinoprb,story2014}.
For example, in contrast to $GW$, the approach explains the
multiple-plasmon satellites observed in x-ray photoemission
spectra (XPS) \cite{guzzo,Aryasetiawan,KRC,ajlee,sky}. However, the behavior in WDM is hitherto unexplored and exhibits a number of unusual and unexpected properties.
As an application relevant to FT DFT, we have implemented the
approach for the homogeneous electron gas (HEG).
Our results show that besides reductions in quasi-particle energy shifts 
(and hence band-gaps) with increasing $T$, the spectral function broadens
and the excited states become strongly damped, corresponding to short
mean-free-paths, blurring of the conventional quasi-particle picture, and smeared-out band-structure.
Finally thermodynamic properties including exchange-correlation
energies are calculated using the
Galitskii-Migdal-Koltun (GMK) sum rule \cite{Martin59,Mahan2000,Koltun},
which serves as a quantitative check on our approximations and yields results
that compare well with existing theories \cite{RPIMC},
and with FT DFT functionals \cite{Trickey14}.


Briefly our approach is based on the finite-$T$ retarded
Green's function formalism \cite{Mahan2000}.
Below we outline the key elements of the approach.
The retarded one-particle FT Green's function $G(\omega)$
satisfies a Dyson equation $G = G^0 + G^0\Sigma G$ \cite{dashgodby},
where $\Sigma$ is the retarded self-energy. 
Formally $\Sigma$ can be found by analytical continuation
of the Matsubara self-energy to real
frequencies, and can be expressed in terms of $G$, the
screened Coulomb interaction $W = \epsilon^{-1} v$, and a
vertex function $\Gamma$ \cite{hedin99review,dashgodby}.
Here and below matrix indices are suppressed unless otherwise
specified, and we use atomic units $e=\hbar=m=1$.
Most practical calculations currently ignore vertex corrections ($\Gamma=1$).
With this restriction, a variety of FT approximations have been used:
For example, the $GW$ approximation is based on the Dyson equation
and MBPT to first order in $W$; then $\Gamma = 1$ and
$G = G^0 + G^0\Sigma^{GW} G$ \cite{hedin99review,hybertsenlouie}.
As a further approximation, the quasi-particle self-consistent $GW$ approach
(QPSCGW) \cite{Schilfgaarde06} starts with the GF on the
Keldysh contour; while self-consistency is carried out,
vertex corrections, satellites, and damping are ignored.
In contrast the retarded cumulant GF method used here is based on an
exponential representation (see below) that builds in implicit
dynamic vertex corrections \cite{hedin99review,guzzo}.
This form can be justified using
the quasi-boson approximation \cite{hedin99review}, in which electron-electron
interactions are represented in terms of electrons coupled to
bosonic excitations.  This approach is an improvement over $GW$ for
spectral properties \cite{sky}, and is exact for certain
models \cite{langreth70}.
More elaborate GF methods exist at least in principle, including
higher order MBPT \cite{gunnarsson1994,Pavlyukh}, and
dynamical mean-field theory with impurity Green's
function's \cite{Deng13,biermann12}, but are more demanding computationally.

The retarded cumulant expansion begins with an exponential ansatz
for the FT GF in the time-domain for a given single-particle state $k$
in which $G$ is assumed to be diagonal, and the spectral function
$A_k(\omega)$ is obtained from its Fourier transform,
\begin{eqnarray}
G_k(t) &=& -i \theta(t) e^{-i\varepsilon^{x}_k t} e^{{\tilde C}_k(t)},\\
A_k(\omega) &=& - \frac{1}{\pi}{\rm Im}\, \int d\omega e^{i\omega t} G_k(t).
\end{eqnarray}
In this formulation static exchange and correlation contributions are
separated $C_k(t) = -i \Sigma_k^{x} t + \tilde C_k(t)$,
and all correlation effects are included in
the dynamic part $\tilde C_{k}(t)$ of the cumulant.
Here $\Sigma^{x}_k = \Sigma_{\bf q}n_{\bf k-q}v_{\bf q}$
is the exchange part of the FT Hartree-Fock one-particle energy,
$\varepsilon_{k}^{x} = \varepsilon_k + \Sigma_{k}^{x}$, 
$\varepsilon_k = k^2/2$ is the bare energy, and $v_{\bf q}=4\pi/q^2$
is the bare Coulomb interaction.
This formulation is directly analogous to that for $T=0$ \cite{KasRC},
apart from implicit temperature dependence in it's ingredients.
The retarded cumulant  $C_k(t)$
can be obtained by matching terms in powers of $W$
to those of the Dyson equation \cite{KasRC,sky}.
Carried to all orders the cumulant GF is formally exact; however, 
by limiting the theory to first order in $W$, $G^0C = G^0\Sigma^{GW} G^0$,
the retarded GW self energy $\Sigma^{GW}$ is sufficient to define
the FT cumulant. $\tilde C(t)$ has a Landau representation,
which implies a positive-definite spectral function and
conserves spectral weight \cite{landau44,langreth70,hedin99review},
\begin{eqnarray}
\label{eq:cum}
{\tilde C}_{k}(t) &=&  \int d\omega \frac{\gamma_{k}(\omega)} {\omega^{2}}
(e^{-i\omega t} + i\omega t -1), \\
\gamma_k(\omega) &=& \frac{1}{\pi}
  \left|{\rm Im}\,\Sigma_{k}(\omega+\varepsilon_{k})\right|.
\end{eqnarray}
The kernel $\gamma_k(\omega)$ reflects the quasi-boson
excitation spectrum in the system, with peaks corresponding to those
in $W(\omega) \propto {\rm Im}\Sigma_k(\omega +\varepsilon_k)$. 
\begin{figure}[t]
  \includegraphics[height=0.85\columnwidth, angle=-90]{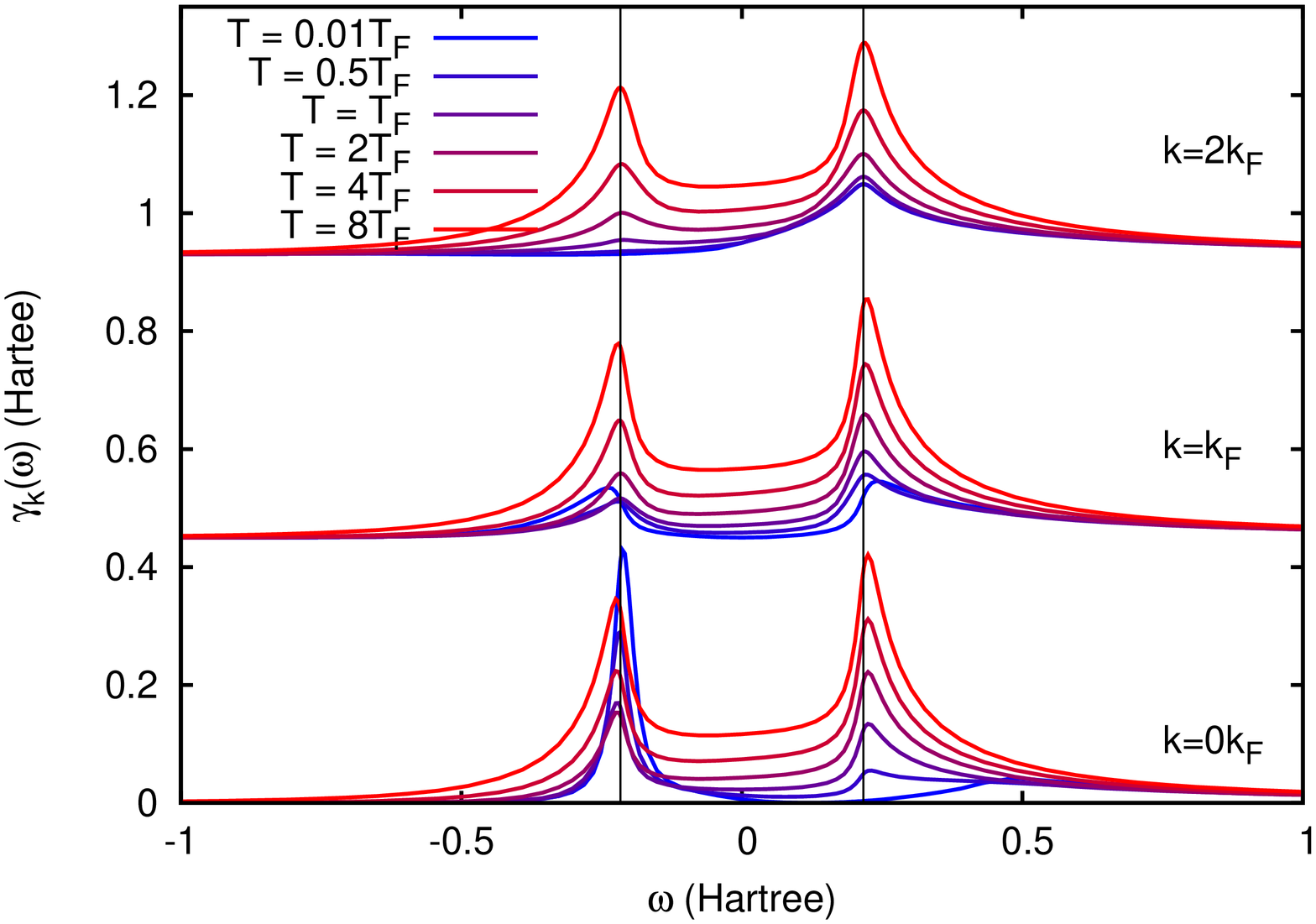}
  \includegraphics[height=0.80\columnwidth, angle=-90]{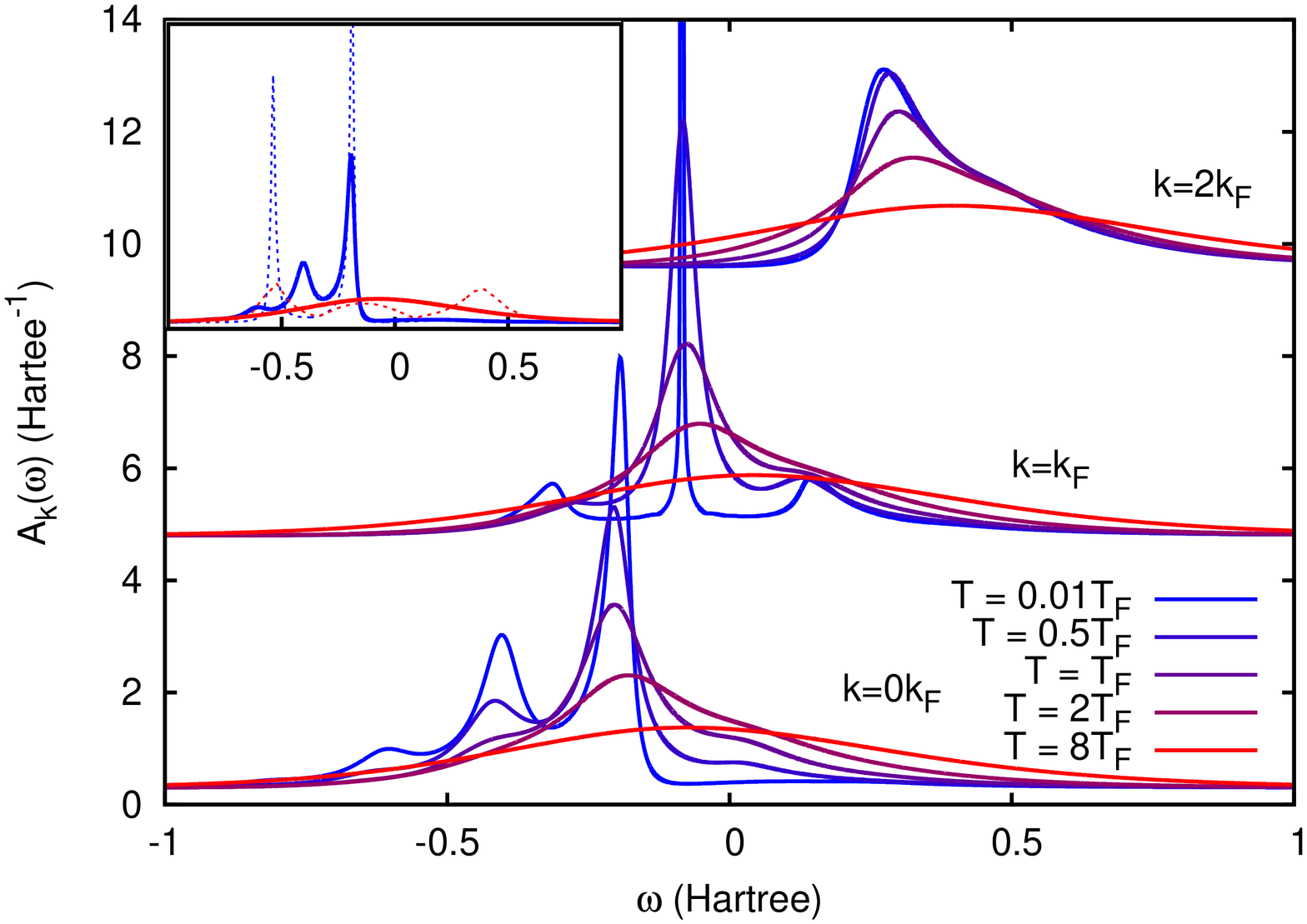}
  \caption{(Color online) FT retarded cumulant kernel $\gamma_k(T,\omega)$ (top);
    and spectral function $A_{k}(T,\omega)$ for the HEG (bottom) 
   for $r_s=4$. Vertical lines in the top plot are shown at $\pm \omega_p$.
    Note the enhanced symmetry of both $\gamma_k$ and $A_k$ at high-$T$. The
inset shows a comparison of the cumulant (solid) and GW (dashes)
spectral functions at $k=0$ for the lowest and highest temperatures. 
\label{fig:spfcn}}
\end{figure}

The basic ingredients in the theory Eq.\ (1-3), are thus $G^0$, 
the $GW$ self-energy $\Sigma^{GW}$,
and the screened Coulomb interaction $W(q,\omega)=\epsilon^{-1}(q,\omega)v_q$,
where $\epsilon(q,\omega)$ is the dielectric function.
These quantities can be calculated using standard
finite-$T$ MBPT \cite{Allen1982,Mahan2000},
starting from the Matsubara Green's function.
The finite-$T$ analog of the $GW$ self energy
for electrons coupled to bosons
is (cf.\ the Migdal approximation \cite{Allen1982}),
\begin{eqnarray}
&&\Sigma^{GW}(\omega,T) = \int d\omega' \frac{d^3 q}{(2\pi)^3}
|{\rm Im}\, W(q,\omega')|\times \nonumber \\  
 && \times \left[ \frac{f(\varepsilon_{k-q}) +
N(\omega')}{\omega+\omega'-\varepsilon_{k-q}+i \delta} + 
\frac{1- f(\varepsilon_{k-q}) +
N(\omega')}{\omega-\omega'-\varepsilon_{k-q}+i \delta}\right].
\end{eqnarray}
Here $N(\omega)=1/(e^{\beta\omega}-1)$ is the Bose factor, 
$f(\varepsilon)=1/(e^{\beta(\varepsilon-\mu)} + 1)$ the Fermi factor,
$\beta=1/k_B T$, and $\mu(T)$ is the chemical potential, as determined below.
The imaginary part of $\Sigma^{GW}(\omega)$ yields
$\gamma_k(\omega)$ (Fig.~\ref{fig:spfcn}~top).  At  high-$T$
the behavior of $\gamma_k$ is dominated
by the Bose factors $N(\omega) \sim k_B T/\omega$ and becomes
strongly symmetric about $\omega=0$.
To obtain $W$ we use the FT-RPA approximation for
the dielectric function,
\begin{equation}
\epsilon(q,\omega) = 1 + 2 v_q \int \frac{d^3 k}{(2\pi)^3}
\frac {f_{k+q} - f_k}{\omega -\varepsilon_{k+q}+\varepsilon_k}.
\end{equation}
The imaginary part of $\epsilon(q,\omega)$ is
analytic \cite{Khanna76,Arista83}, and the real part is  
calculated via Kramers-Kronig transform. This yields
the finite-$T$ loss function
$L(q,\omega) = |{\rm Im}\, \epsilon^{-1}(q,\omega)|$.
For the HEG $L(q>0, \omega)$ exhibits broadened and blue-shifted
plasmon-peaks with increasing $T$ \cite{Arista83}.
The chemical potential $\mu=\mu(T,N)$ implicit in the Fermi factors
$f(\omega)$ is determined
by enforcing charge conservation $\Sigma_{\bf k}\, n_k = \langle N(T)\rangle$,
where the single-particle occupation numbers $n_k(T)$
are given by a trace over the spectral function \cite{Martin59},
\begin{equation}
  n_k(\mu,T) = \int_{-\infty}^{\infty} d\omega\, A_k(\omega)f(\omega), 
\end{equation}
Occupation numbers can be measured e.g., by Compton
scattering \cite{Seidler12,klevak2014,Huotari},
and are sensitive to the many-body correlation effects in $A_k$.

Note that  at low-$T$, $A_k(\omega)$ exhibits multiple-satellites for $k<k_F$
while for $k>k_F$ and at $T>T_{F}$, the quasi-particle peak broadens,
and overlaps the satellites. Thus in WDM the structure of $A_k$ blurs into
single asymmetric peak with a centroid at $\varepsilon_{k}^{x}(T)$
and root mean square width $\delta_k$ given by the 2nd cumulant moment of $A_k(\omega)$
$\delta_k^2=\tilde C_k''(0) = \int d\omega\, \gamma_k(\omega)$.
A dimensionless measure of correlation is given by the
satellite strength in the spectral function $a_k\equiv \ln (1/Z_k) =
\int d\omega \gamma_k(\omega)/\omega^2$,
where $Z_k$ is the renormalization constant, which is determined from
the last term in Eq.\ (\ref{eq:cum}).
 This measure corresponds to the mean number
of bosonic  excitations and is of order $0.2 r_s^{3/4}$ ($r_s$
being the Wigner-Seitz radius) for plasmons at $T=0$ \cite{hedin99review}.
Surprisingly $a_k$ is only weakly dependent on temperature
with $Z_{k_F}\approx0.6$ at $r_s=4$.
Formally the structure of the Landau cumulant in Eq.\ (\ref{eq:cum}) is
consistent with the conventional quasi-particle picture, i.e.,
a renormalized main peak red-shifted by a ``relaxation energy" $\Delta_k$
and a series of satellites. 
The correlation part of the quasi-particle energy shift
$\Delta_k$ is obtained from middle term in Eq.\ (\ref{eq:cum}),
while the first term gives rise to
satellites at multiples of the plasmon peak $\omega_{p}$.
The quasi-particle energy is then
$\varepsilon^{qp}_k = \varepsilon_k  + \Delta_k$,
\begin{figure}[t]
    \label{fig:delta}
\includegraphics[height=0.80\columnwidth, angle=-90]{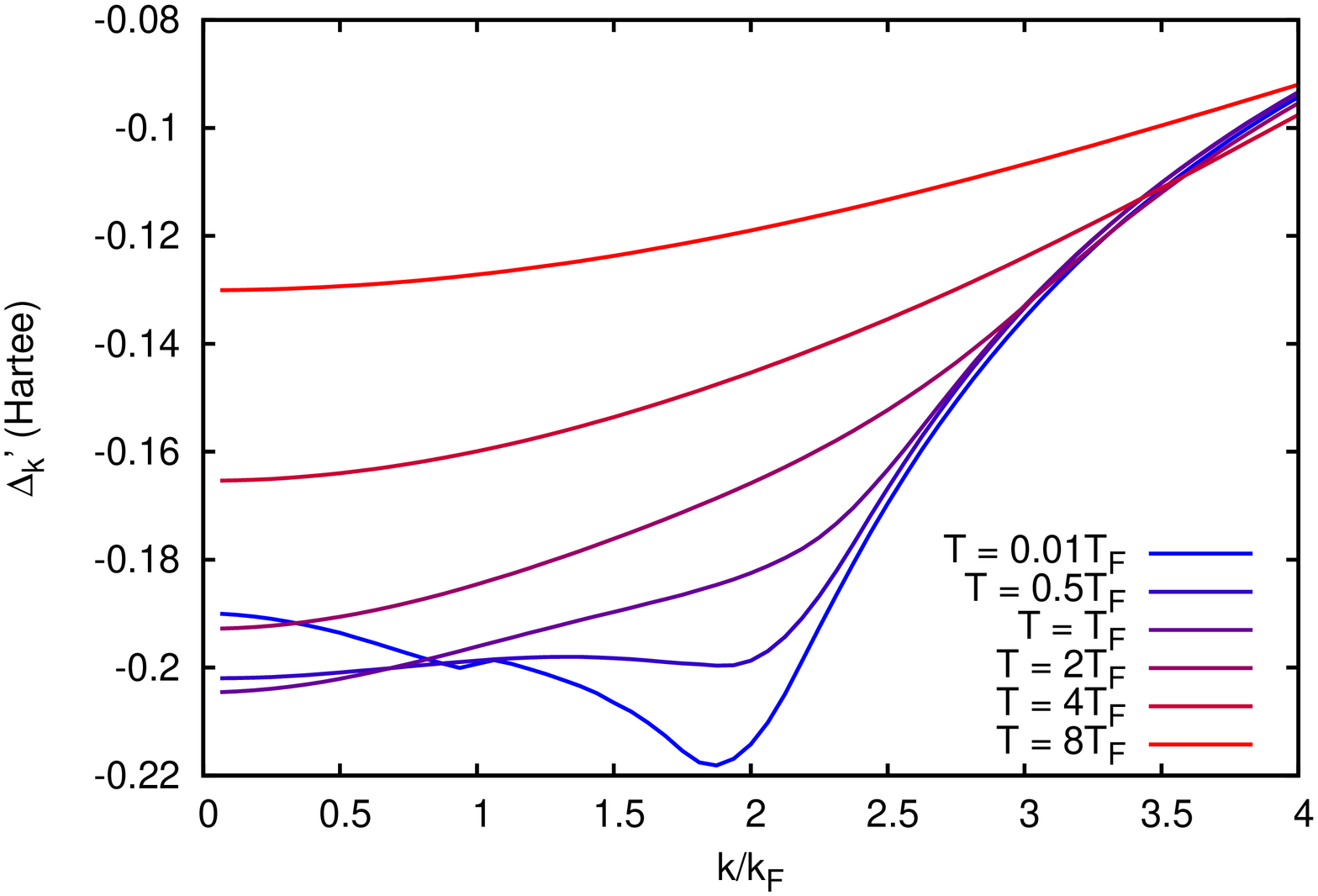}
\includegraphics[height=0.80\columnwidth, angle=-90]{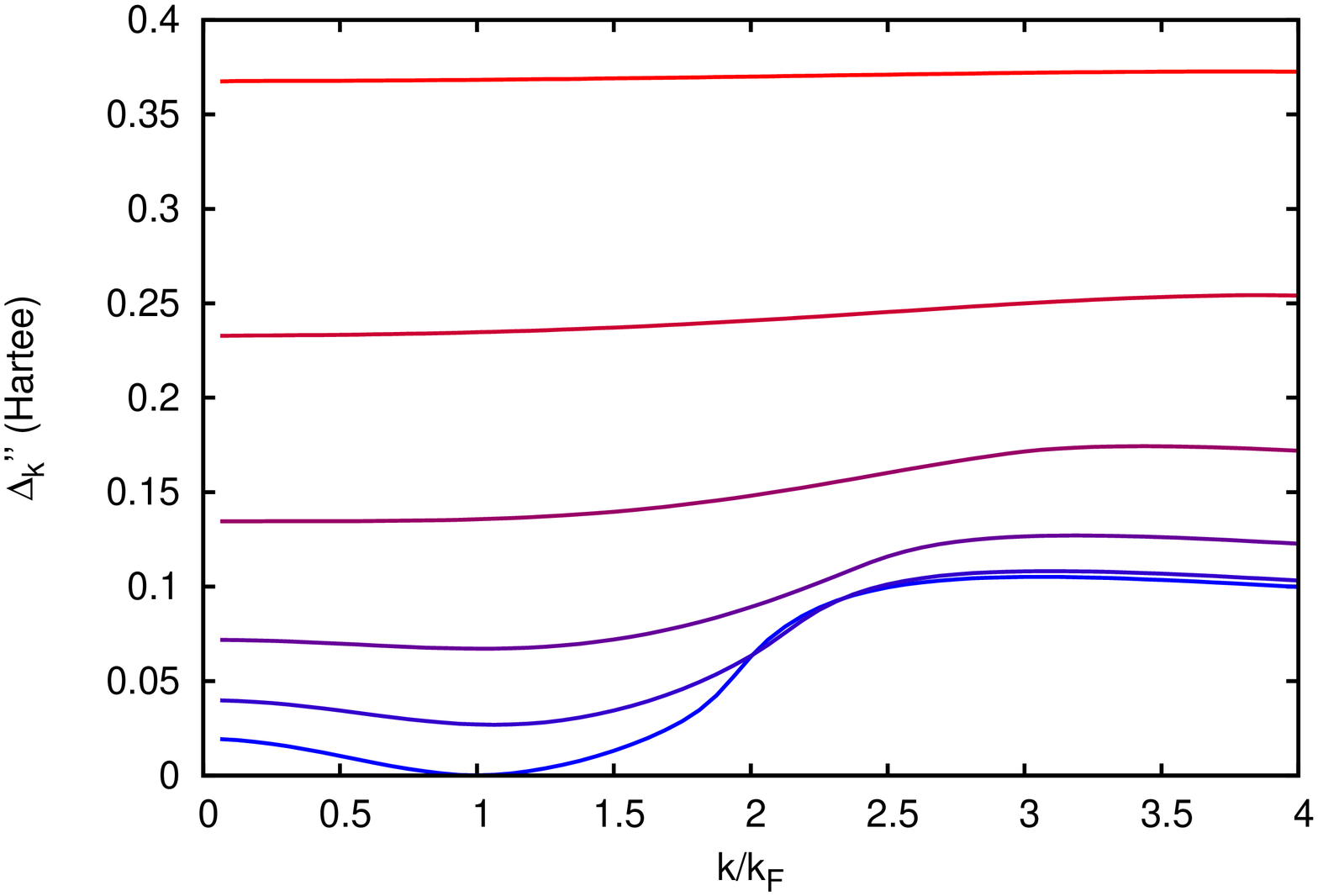}
\caption{(Color online) Real (top) and imaginary (bottom) parts of the quasiparticle correction $\Delta_{k}$ shown for varying temperature.}
\end{figure}
where 
\begin{equation}
\Delta_k = \Sigma_{k}^{x} + \int d\omega\, \frac{\gamma_{k}(\omega)}
      {(\omega-i\delta)}.
\end{equation}
The real part $\Delta_k' $ is the relaxation energy which is 
comparable to that in QPSCGW \cite{Schilfgaarde06}.
Due to the increasingly symmetrical
behavior of $\gamma_k(\omega)$, $\Delta'$ decreases smoothly with $T$.
However, a striking difference with the $T=0$ behavior is the presence of
an imaginary part $\Delta_k''$ even at the Fermi momentum, which
becomes large at high-$T$ since $\gamma_k(0)\neq 0$. This behavior implies strongly damped
propagators that blur the usual quasi-particle picture,
smearing band-gaps and band-structures. This behavior is clearly
evident in the spectral function 
$A_k(\omega)=(1/\pi)|{\rm Im}\, G_k(\omega)|$ \cite{Mahan2000,Martin59},
which is directly related to x-ray photoemission spectra (XPS)
(Fig.\ \ref{fig:spfcn}). 
In contrast, the GW spectral function retains satellite
structure even at high $T$ (Fig.~\ref{fig:spfcn} inset).

\begin{figure}[t]
  \includegraphics[height=0.80\columnwidth, angle=-90]{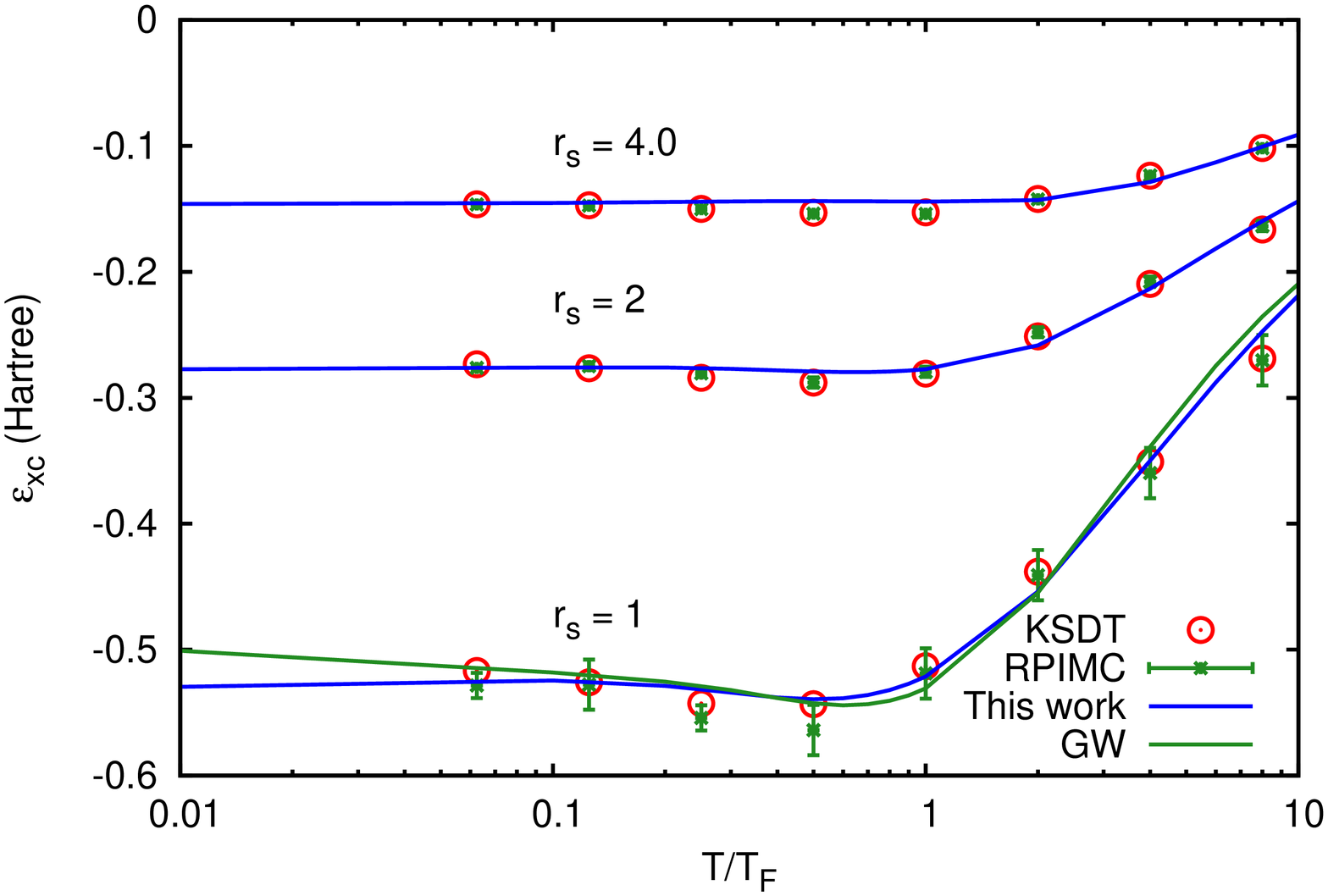}
  \includegraphics[height=0.80\columnwidth, angle=-90]{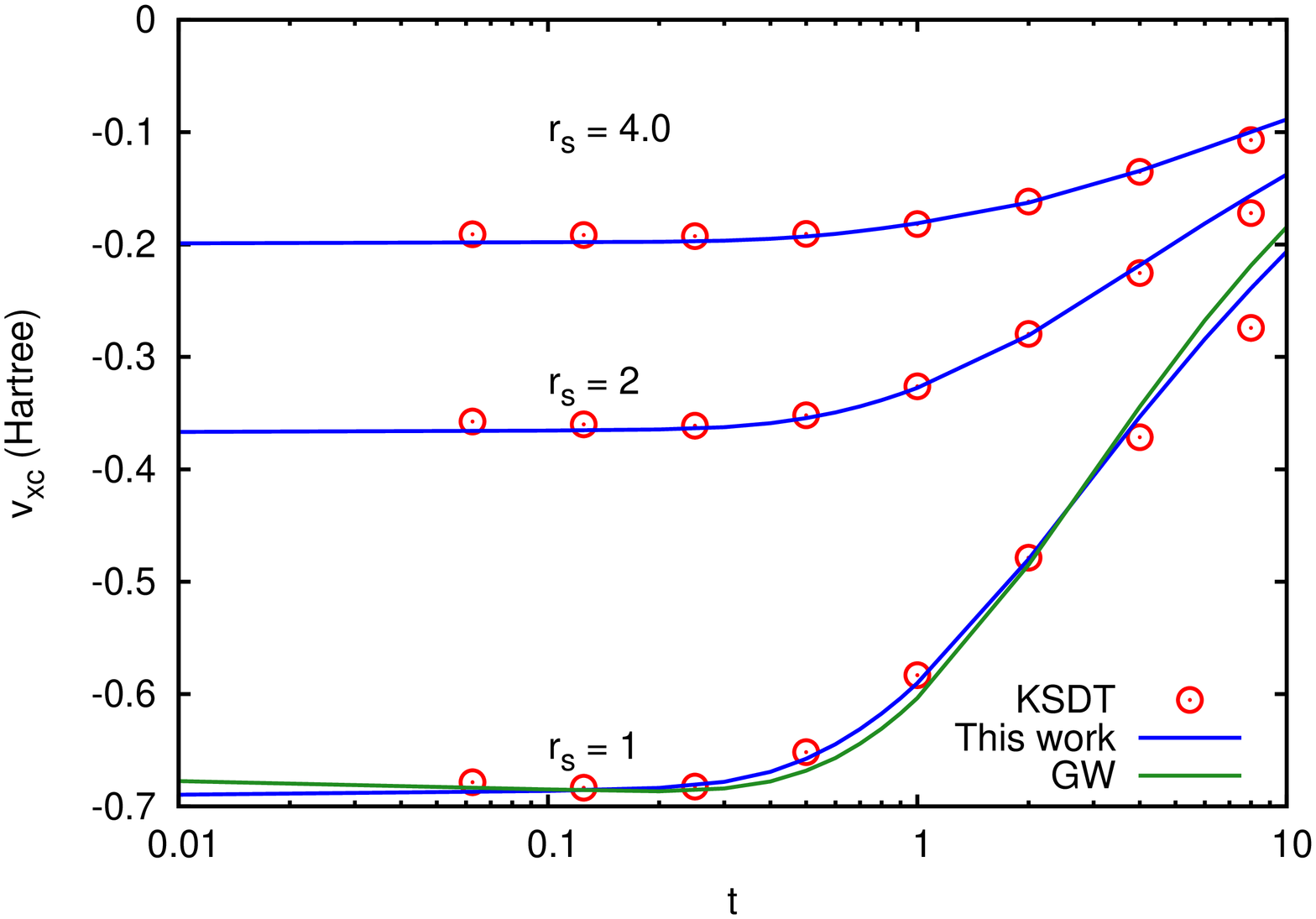}
  \caption{(Color online) Finite-$T$ exchange-correlation energy per particle (top) and potential $v_{xc}$ vs $\tau=T/T_F$ for
    the HEG from the cumulant expansion (blue), compared to
    RPIMC (crosses) and constrained fits (red circles). The inset shows the quasiparticle renormalization factor at the Fermi momentum $Z_{k_F}$ as a function of $T$.\label{fig:exc}}
\end{figure}

One of the advantages of the cumulant formalism is that is provides an alternative approach 
to calculate thermodynamic equilibrium properties.
Remarkably, knowledge of $\mu(T)$ is sufficient to determine the
FT DFT exchange-correlation potential for the HEG \cite{Dharma2000},
$v_{xc}(T)\equiv\mu_{xc}(T)=\mu(T) - \mu_{0}(T)$, where $\mu_0(T)$ is
the chemical potential for non-interacting electrons (Fig.\ \ref{fig:exc}).
Clearly the agreement with existing FT DFT exchange-correlation
potentials and theoretical calculations is quite good.
Moreover, the FT total energy per particle $\varepsilon(T) \equiv
E(T)/N$,
can be calculated from the GMK
sum-rule \cite{Martin59,Mahan2000,Koltun}, 
\begin{align}
  \label{eq:gmenergy}
  \varepsilon(T) &= \sum_{k} \int d\omega \,
  \left[ \omega + \varepsilon_k  \right] A_k(\omega)f(\omega) ,
  \nonumber \\
  &=\varepsilon_H + \varepsilon_{xc},
\end{align}
which is valid for any
Hamiltonian with only pair interactions. 
Here $\varepsilon_H$ is the Hartree energy.
This relation is similar in form to the zero-$T$ Galitskii-Migdal sum-rule,
except for the Fermi factor.
Our results for
$\varepsilon_{xc}(T)$ for the HEG are shown in Fig.~\ref{fig:exc}.
\begin{figure}[t]
\includegraphics[height=0.80\columnwidth, angle=-90]{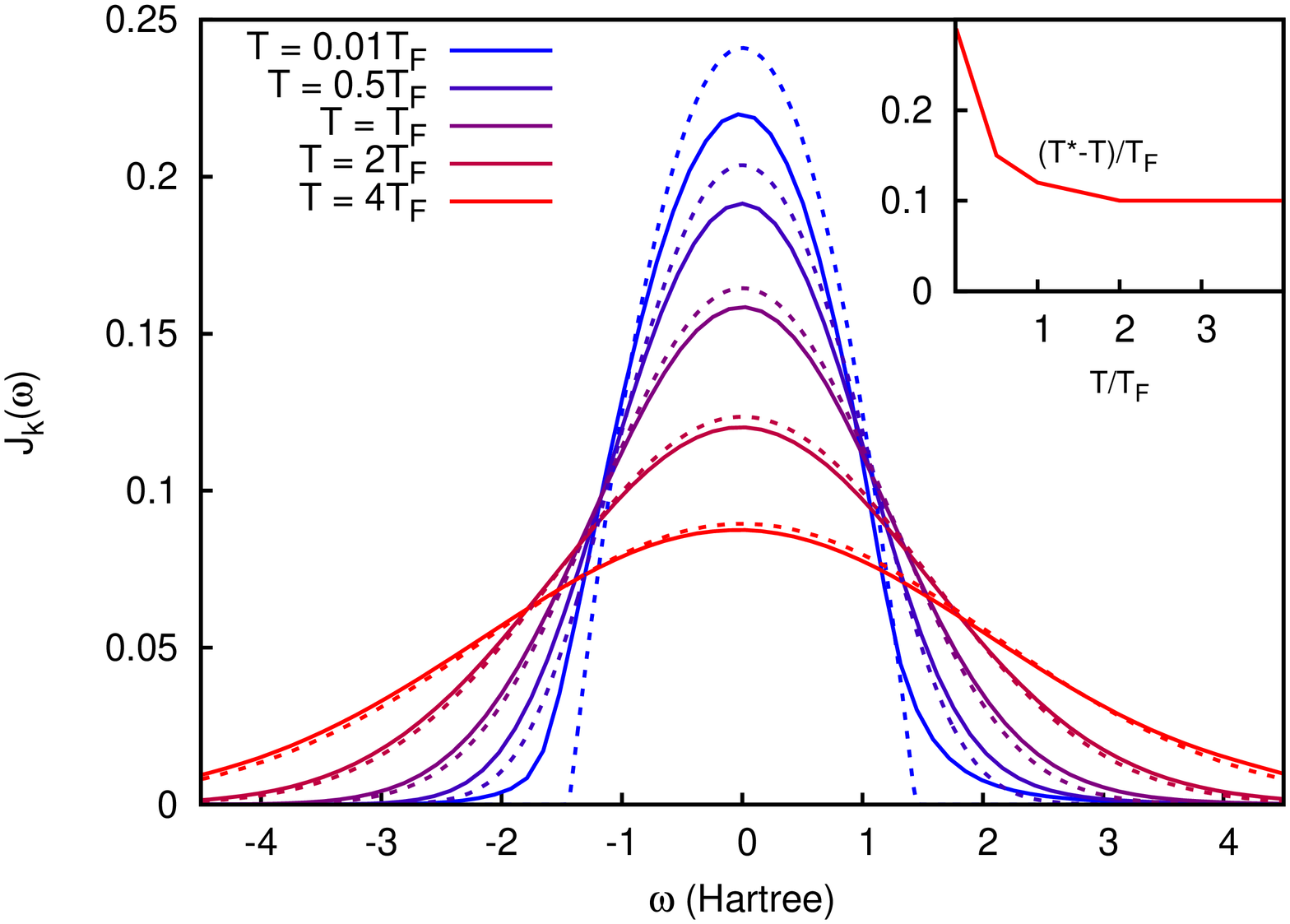}
\caption{(Color  online) Finite-$T$ Compton spectrum 
  compared to the free-electron $T=0$ result at $r_s=4$ and
  ${q} = 3$ Bohr$^{-1}$.
  The inset
  shows the difference between the effective temperature $T^*$ at which
  the free-electron calculation matches the
  cumulant result at $T$.\label{fig:compton}}
\end{figure}
Clearly the agreement between the cumulant results, accurate restricted path-integral Monte Carlo (RPIMC)
calculations \cite{RPIMC}, and existing
FT DFT functionals \cite{Trickey14} is quite good.
At $T=0$, $\varepsilon_{xc}$  was found to be slightly better
with the retarded cumulant than with $G_{0}W_{0}$, 
but self-consistent $GW$ gave better total energies \cite{KasRC}.
Finally, we calculate the Compton spectrum (Fig.\ \ref{fig:compton}) following Ref.\ \cite{Schulke},
\begin{equation}
  J_{q}(\omega) = \sum_{\bf
  q}\int d\omega A_{\bf k}(\omega)A_{\bf
    k+q}(\omega+\omega')f(\omega)f(\omega+\omega').
\end{equation}
At small $T$, effects of correlation are very noticeable, leading to an effective
temperature $T^*$ (i.e. the temperature at which free-electron calculations
match the interacting one) of $(T^*-T)/T_F \approx 0.3$ (see
inset), while at high-$T$  the effect is smaller but non-negligible,
$(T^*-T)/T_F\approx 0.1$.


 In summary we have developed a finite-$T$ Green's function approach
for calculations of both excited state and thermodynamic properties
over a wide range of densities and temperatures.
 Our approach is based on the retarded cumulant expansion
to first order in $W$.
This approximation greatly simplifies the theory and provides a practical
approach both for calculations and the interpretation of exchange and
correlation effects in terms of the behavior of retarded
cumulant $C_k(t)$, which is directly related to the FT GW self energy
$\Sigma^{GW}$.
The cumulant GF builds in an approximate dynamic vertex, going beyond
the $GW$ approximation, yet is no more difficult to calculate.  
Differences with respect to GW provide a measure of vertex
effects and hence the accuracy of the theory. 
 Although we have focused on the HEG, reflecting the importance 
of density fluctuations at high-$T$, the cumulant can be generalized
to include additional quasi-boson excitations such as phonons since
the leading cumulant 
is linear in bosonic couplings \cite{story2014}.
The method provides an attractive complement to FT DFT, RPA, and
RPIMC methods which are appropriate for correlation energies but inapplicable
for many excited state properties.  Illustrative results for the HEG explain
the crossover in behavior from cool to WDM regimes and
the blurring of the conventional quasi-particle picture. The crossover
is largely due to an enhanced coupling to density fluctuations at high-$T$.
We find that correlation effects remain strong at all temperatures
and can have significant effects on spectral properties.
Calculations of thermodynamic equilibrium quantities including
exchange-correlation energies and potentials are in good agreement
- typically within a few percent -
 with existing DFT functionals and quantum Monte Carlo (QMC) calculations. 
Many extensions are possible, ranging from
excited state, spectroscopic to thermodynamic properties of realistic
systems, and potentially to the development of improved FT DFT
functionals \cite{Burke16} e.g., in regimes inaccessible to conventional methods.

Acknowledgments: We thank K. Burke, G. Bertsch, V. Karasiev, L. Reining,
E. Shirley, G. Seidler, T. Devereaux, and S. Trickey for comments and suggestions. 
This work is supported by DOE BES Grant DE-FG02-97ER45623.

\end{document}